\documentstyle[12pt]{article}
\begin{document}
\rightline{LANDAU-93-04/GR-1}
\rightline{April 1993}
\vspace{1cm}
\centerline{\bf \large NEW RESTRICTIONS ON SPATIAL TOPOLOGY OF THE UNIVERSE}
\centerline{\bf \large FROM MICROWAVE BACKGROUND TEMPERATURE FLUCTUATIONS}
\vspace{0.5cm}
\centerline{\large A. A. Starobinsky}
\vspace{0.5cm}
\centerline{\it L. D. Landau Institute for Theoretical Physics, Russian
Academy of Sciences}
\centerline{\it 117334, Moscow, Russia}
\vspace{0.5cm}
\centerline{Submitted to {\it JETP Letters} 15 April 1993}
\vspace{1cm}
If the Universe has the topology of a 3-torus ($T^3$), then it follows
from recent data on large-scale temperature fluctuations $\Delta T/T$
of the cosmic microwave background that either the minimal size of the
torus is at least of the order of the present cosmological horizon, or the
large-scale $\Delta T/T$ pattern should have a symmetry plane (in case of
the effective $T^1$ topology) or a symmetry axis (in case of the effective
$T^2$ topology), the latter possibility being probably excluded by the data.
\vspace{1.5cm}

The possibility that a spacelike hypersurface $t=const$ (or
$\varepsilon=const$)
of the Friedmann-Robertson-Walker (FRW) cosmological model may possess a
non-trivial (i.e. not $R^3$) topology was known long ago in connection with
the $S^3$ topology in the case of a closed Universe
($\Omega_{tot}\equiv \varepsilon_{tot}/\varepsilon_{crit}>1$). In this case,
it follows from the age of the Universe and direct dynamical estimates of
matter density that the corresponding topological size - the radius of the
Universe - cannot be significantly less than the present cosmological horizon.
A way to get small topological lengths is to assume a spatially-flat 3-space
with the 3-torus topology ($T^3$), i.e. to make the following identifications
of points of the 3-D flat hypersurface $t=const$:
\begin{equation}
x\equiv x+L_1,\; y\equiv y+L_2,\; z\equiv z+L_3, \   \ L_1\geq L_2\geq L_3\,.
\end{equation}
This possibility was considered in $^{1-3}$ with the conclusion that
$aL_1$ should exceed $500-1000$ Mpc where $a\equiv a(t_0)$ is the present
scale factor of the Universe (in this paper, all scales are given
for $H=50$ km/s/Mpc). If $aL_1\gg R_{hor}$ where $R_{hor}$ is the present
size of the cosmologocal horizon ($R\approx 12000$ Mpc if
$a(t)\propto t^{2/3}$ now), then the effective spatial topology of the
Universe becomes $T^2$;
for $aL_2\gg R_{hor}$, it is effectively $T^1$. Note also that the remaining
possibility to have a non-trivial topology with small characteristic sizes
in an open FRW Universe $^4$ is practically excluded because it requires
$\Omega_{tot}\ll 1$ that contradicts observational data ($\Omega_m=0.2-0.4$
from dynamical estimates based on optical galaxies; use of IRAS galaxies
yields a larger value for $\Omega_m$).

Classical general relativity does not put any restrictions on $L_i$, they
are just a part of initial conditions for the Universe at the singular
hypersurface $a=t=0$. Different versions of the cosmological scenario with
a de Sitter (inflationary) phase in the Early Universe $^5$ do not preclude
this topology, too. In this case, the $T^3$ topology may be a result of
quantum creation of the Universe $^6$ (if this process is believed to be
properly described by a WKB solution of the Wheeler-De Witt equation
exponentially damping under a potential barrier in the direction from
$a=0$ to the border of a classically allowed region). Then, however, as a
result of inflation, the present topological sizes $aL_i$ are
expected to be much more than $R_{hor}$ with an overwhelming probability.
It is also possible to generate a non-trivial topology of the
$\varepsilon=const$ hypersurface during inflation $^7$ but not of the $T^3$
type. Thus, the prediction of the inflationary scenario is that spatial
topology of the Universe should be trivial at all observable scales up to
$R_{hor}$ and even larger.

Irrespective of any assumptions of the Early Universe behaviour, it becomes
possible now to strongly restrict the hypothesis of the spatial $T^3$
topology of the Universe directly from recent observational data $^{8,9}$ on
large-scale fluctuations of the cosmic microwave background radiation
\footnote {Results of this paper were presented in the author's
plenary talk at the Texas/PASCOS'92 conference (Berkeley, USA, Dec. 13-18,
1992). After that the author became aware of recent papers $^{10}$ where
similar results for the $L_1=L_2=L_3$ case were obtained independently.}$^)$.
Here, the COBE data $^9$ appear to be the most appropriate. If
$\Delta T/T$ fluctuations are produced by adiabatic perturbations and
$a(t)\propto t^{2/3}$ since decoupling till the present time
(the matter-dominated regime in a spatially flat FRW Universe), then the
Sachs-Wolfe formula $^{11}$ is valid for large angles
$\vartheta\geq 2^{\circ}$ ($l\leq 30$):
\begin{equation}
{\Delta T\over T}(\theta ,\varphi)={1\over 3}\Phi (R_{rec}/a,\theta ,\varphi)=
-{1\over 10}h(R_{rec}/a,\theta ,\varphi)
\end{equation}
where $\Phi (\vec r)$ - the gravitational potential - defines a perturbed
space-time metric in the longitudinal gauge:
\begin{equation}
ds^2=(1+2\Phi)dt^2-a^2(t)(1-2\Phi)(dx^2+dy^2+dz^2)\, ,
\end{equation}
$h(\vec r)$ is a metric perturbation in the ultra-synchronous gauge
(where $h_{\alpha}^{\beta}\approx h(\vec r)\delta_{\alpha}^{\beta}$ as
$t\to 0$, $\alpha ,\beta =1,2,3$) and $R_{rec}$ is the present
radius of the surface of last scattering, $R_{rec}\approx 0.97R_{hor}$.
$\Phi$ and $h$ do not depend on $t$ in linear approximation.

In a 3-torus Universe, the spectrum of Fourier modes is discrete:
\begin{eqnarray}
\Phi(\vec r)=\sum_{npq} \Phi(\vec k)e^{i\vec k\vec r}, \   \
\vec k=\left({2\pi n\over L_1},\, {2\pi p\over L_2},\,
{2\pi q\over L_3}\right), \nonumber \\
n,p,q\in {\bf Z}, \  \ n^2+p^2+q^2\not= 0
\end{eqnarray}
(the $n=p=q=0$ mode is a gauge one). $\Phi(\vec k)$ is not assumed to have
the flat (Harrison-Zeldovich) spectrum (contrary to $^{10}$). Expanding
$e^{i\vec k \vec r}$ in terms of normalized spherical harmonics, we get:
\begin{eqnarray}
{\Delta T\over T}=\sum_{lm} \left({\Delta T\over T}\right)_{lm}
Y_{lm}(\theta ,\varphi), \  \
\left({\Delta T\over T}\right)_{lm}={4\pi\over 3}\sum_{npq}i^lj_l(kR_{rec}/a)
Y_{lm}^{\ast}(\vec k/k), \nonumber \\
k=|\vec k|, \  \ j_l(v)=\sqrt{{\pi\over 2v}} J_{l+1/2}(v).
\end{eqnarray}
Total multipole amplitudes $(\Delta T/T)_l$ follow from
the expression
\begin{eqnarray}
\left({\Delta T\over T}\right)_l^2\equiv{1\over 4\pi}\sum_{m=-l}^l
\left({\Delta T\over T}\right)_{lm}^2= \nonumber \\
{2l+1\over 9}\sum_{\vec k}\sum_{\vec k'}\Phi(\vec k)\Phi(\vec k')
j_l(kR_{rec}/a)j_l(k'R_{rec}/a)P_l\left({\vec k\vec k'\over kk'}\right).
\end{eqnarray}

Let $aL_1\ll R_{hor}$ (a small 3-torus Universe). Then
$\left({\Delta T\over T}\right)_l^2\propto (2l+1)$ (this result was already
obtained in $^{12}$). In particular, for a stochastic isotropic
$\delta$-correlated spectrum of perturbations
$\langle \Phi(\vec k)\Phi(\vec k')\rangle=\Phi^2(k)\delta_{\vec k \vec k'}$,
\,Eq.(6) takes the form:
\begin{equation}
\left({\Delta T\over T}\right)_l^2={2l+1\over 9}\sum_{\vec k}
\Phi^2(k)j_l^2(kR_{rec}/a)\approx {2l+1\over 18}\sum_{\vec k}
\Phi^2(k)\left({a\over kR_{rec}}\right)^2.
\end{equation}
Therefore, amplitudes of low multipoles are much smaller (at least, in
$aL_1/R_{hor}$ times) than $\Delta T/T$ fluctuations at angles
$\vartheta \sim aL_1/R_{hor}$ (in radians) \footnote {Here, the author would
like to rectify an incorrect statement that $(\Delta T/T)_2=0$ in the case
$L_1=L_2=L_3$ that appeared in the beginning of Ref.6 (fortunately,
it was not used in any way in the following body of the paper). The correct
statement is that this quantity is small (much less than $10^{-5}$) if
$aL_1\ll R_{hor}$.}$^)$.

Multipole dependence of the observed large-scale $\Delta T/T$ fluctuations
$^9$ does not follow this relation. Actually, it is much better fitted by the
law $(\Delta T/T)_l^2\propto {2l+1\over l(l+1)}$ following from the
inflationary scenario $^{13}$. Thus, a sufficiently small 3-torus Universe
is excluded by the data. Let us find more exact restrictions on the model.
Another possible fit to the angular correlation function $\xi_T(\vartheta)$
of the COBE data is a Gaussian with the correlation angle
$\vartheta_c=13.5\pm2.5^{\circ}$ (see $^{14}$). This correspond to $l_m=3-5$
where $l_m$ is the multipole with the largest amplitude $(\Delta T/T)_l$.
So, we may assume that $l_m\leq 6$.

First, consider modes with $npq\not= 0$ having a generic dependence on
$\vec r$ and ($\theta ,\varphi$). For them, $k>2\pi /L_3$. The quantity
$(2l+1)j_l^2(v)$ considered as a function of integer $l$ has a maximum
at $l=l_m\leq 6$ for $v<8.5$. Thus, a lower limit on the minimal topological
size of the Universe follows:
\begin{equation}
aL_3>0.75R_{hor}\approx 9000\, Mpc.
\end{equation}

Second, from modes with only one of the numbers $n,p,q$ equal to zero,
the modes with $q=0,\,n^2+p^2\not= 0$ are the most important. If the
contribution to $\Delta T/T$ from these modes dominates at low $l$, then
$L_3$ may be much less than $R_{hor}/a$ and the estimate (8) refers to the
intermediate size $aL_2$. However, in this case the large-scale $\Delta T/T$
pattern has a symmetry plane due to the absence of dependence of the
corresponding part of $\Phi(\vec r)$ on $z$, i.e.
\begin{equation}
{\Delta T\over T}(\theta ,\varphi)={\Delta T\over T}(\pi -\theta ,\varphi)
\end{equation}
in some system of angular coordinates. For $aL_2\gg R_{hor}$, the observed
part of the Universe has the effective spatial topology $T^1$.

Finally, from modes with two of the numbers $n,p,q$ equal to zero, the modes
with $p=q=0,\,n\not= 0$ are the most interesting. They may give the main
contribution to $\Delta T/T$ at low $l$ if both $L_2$ and $L_3$ are much
less than $R_{hor}/a$. Then the lower limit (8) refers to the maximal size
$aL_1$ only  but the large-scale $\Delta T/T$ pattern has a symmetry axis:
\begin{equation}
{\Delta T\over T}(\theta ,\varphi)={\Delta T\over T}(\theta).
\end {equation}
For $aL_1\gg R_{hor}$, we come to the case of the effective $T^2$ spatial
topology.

Therefore, three possibilities exist: \\
1) all sizes of a 3-torus satisfy the estimate (8); \\
2) the estimate (8) refers to the two larger sizes but the large-scale
$\Delta T/T$ pattern has a symmetry plane (9); \\
3) the maximal size of the torus alone satisfies (8) and Eq.(10) is valid. \\
The $\Delta T/T$ map obtained in $^9$ is noisy, so it does not immediately
provides the real picture of primordial fluctuations. Nevertheless, Eq.(10)
imposes so much symmetry that the third case is probably excluded already.
The symmetry imposed by Eq.(9) in the second case is less evident visually,
so more accurate maps are required to exclude this case definitely.
If this is achieved, we may be sure that all topological sizes are
at least of the order of the present cosmological horizon - in accordance
with the
prediction of the inflationary scenario. $\Delta T/T$ fluctuations produced
by primordial gravitational waves are similar to those produced by adiabatic
perturbations (though they depend on values of gravitational waves at all
times between $t_{rec}$ and $t_0$), so the same conclusions are expected to
be valid in that case, too.

The financial support for this research was provided by the Russian Foundation
for Fundamental Research, project code 93-02-3631.

\vspace{1cm}
\centerline{\bf REFERENCES}
\vspace{1cm}
\noindent
$^1$Ya. B. Zeldovich, Comm. Astrophys. Space Phys. {\bf 5}, 169 (1973). \\
$^2$D. D. Sokolov and V. F. Shwartsman, Zh. Eksp. Teor. Fiz. {\bf 66}, 412
(1974). \\
$^3$L. Z. Fang and H. Sato, Comm. Theor. Phys. {\bf 2}, 1055 (1983). \\
$^4$D. D. Sokolov and A. A. Starobinsky, Astron. Zh. {\bf 52}, 1041 (1975). \\
$^5$A. A. Starobinsky, Phys. Lett. B {\bf 91}, 99 (1980); A. Guth, Phys. Rev.
D {\bf 23}, 347 (1981); A. D. Linde, Phys. Lett. B {\bf 108}, 389 (1982). \\
$^6$Ya. B. Zeldovich and A. A. Starobinsky, Pisma Astron. Zh. {\bf 10}, 323
(1984). \\
$^7$A. A. Starobinsky, in: Proc. VII Sov. Grav. Conf. (18-21 Oct. 1988),
Erevan Univ. Press, Erevan (1988), p. 455.  \\
$^8$I. A. Strukov, A. A. Brukhanov, D. P. Skulachev and M. V. Sazhin,
Pisma Astron. Zh. {\bf 13}, 65 (1992); Mon. Not. Roy. Astron. Soc. {\bf 258},
37 (1992). \\
$^9$ G. F. Smoot, C. L. Bennett, A. Kogut, E. L. Wright et al., Astroph. J.
{\bf 396}, L1 (1992). \\
$^{10}$I. Yu. Sokolov, Pisma Zh. Eksp. Teor. Fiz. {\bf 57}, this issue (1993);
D. Stevens, D. Scott and J. Silk, preprint (1993).  \\
$^{11}$R. K. Sachs and A. W. Wolfe, Astroph. J. {\bf 147}, 73 (1967). \\
$^{12}$L. Z. Fang and M. Houjun, Modern Phys. Lett. A {\bf 2}, 229 (1987). \\
$^{13}$P. J. E. Peebles, Astroph. J. {\bf 263}, L1 (1982); A. A. Starobinsky,
Pisma Astron. Zh. {\bf 9}, 579 (1983). \\
$^{14}$E. L. Wright, S. S. Meyer, C. L. Bennett, N. W. Boggess et al.,
Astroph. J. {\bf 396}, L13 (1992).

\vfill
\end{document}